\documentclass[10pt]{iopart}

\usepackage{amssymb}

\usepackage{graphicx}

\usepackage{array}
\usepackage[dvipsnames]{xcolor} 

\definecolor{oceanblue}{rgb}{0.0, 0.447, 0.765}
\begin{document}

\title{Frustration and chirality in three-dimensional trillium lattices:
	Insights and Perspectives}

\author{J. Khatua$^{\dagger}$ and Kwang-Yong Choi$^\ast$}

\address{Department of Physics, Sungkyunkwan University, Suwon 16419, Republic of Korea}

\ead{$^\dagger$khatuajoydev47@gmail.com}
\ead{$^\ast$choisky99@skku.edu}

\vspace{2pc}
\begin{indented}
	\item[] \today 
\end{indented}

\begin{abstract}
Condensed matter physics continues to seek new frustrated quantum materials that not only deepen our understanding of fundamental physical phenomena but also hold promise for transformative technologies. In this review article, we highlight the unique features of chiral spin topology and review the topological phenomena recently identified in trillium lattice compounds. Based on the unique spin states realized in these systems, we explore the potential for realizing various theoretically proposed chiral quantum phases. We examine representative materials- including the magnetic insulating compound K$_{2}$Ni$_{2}$(SO$_{4}$)$_{3}$ and  and the intermetallic EuPtSi—discussing both experimental findings and theoretical predictions, while outlining several key questions. Finally, we offer a perspective on promising research directions aimed at uncovering novel emergent behavior in chiral trillium lattice-based materials.
\end{abstract}

\maketitle

\ioptwocol
\section{Introduction}
\subsection{Frustrated quantum magnets}
Frustrated magnets—where symmetry, topology, and quantum fluctuations intertwine—have emerged as a fertile platform for exploring emergent  many-body phenomena in contemporary condensed matter physics~\cite{Balents2010,Savary_2016}.
In these systems, competing interactions inhibit the establishment of a unique ground state~\cite{10.1007978-3-642-10589-0,Bramwell2009}. For instance, in triangular lattices,  antiferromagnetic interactions between nearest-neighbor Ising spins make it impossible to satisfy all pairwise alignments simultaneously, leading to strong spin frustration~\cite{Ramirez2003,PhysRev.79.357}. Beyond geometric constraints, triangular lattices can also host kinetic frustration, a purely quantum mechanical form of spin frustration. In this scenario, although electrons seek to minimize their kinetic energy (approximately $-z|t|$, where $z$ is the coordination number and $t$ is the hopping amplitude), destructive quantum interference between different hopping paths within the lattice prevents this kinetic energy minimization~\cite{Yin2011,PhysRevLett.105.187002}.    \\
While conventional magnets undergo symmetry-breaking phase transitions upon cooling~\cite{landau1937theory,sachdev1999quantum}, highly frustrated systems can evade such transitions, giving rise to exotic quantum states such as spin liquids~\cite{scienceaay0668,ANDERSON1973153}, spin nematics~\cite{podolsky2005properties}, and spin ice~\cite{Bramwell2009}—that lie beyond Landau’s paradigm.
  Furthermore, these states often host fractionalized excitations, where the elementary excitations behave as fractions of the bare electronic degrees of freedom~\cite{PhysRevLett.50.1395}.  In parallel with high-energy physics, such fractionalized excitations can be treated as matter fields coupled to emergent gauge fields~\cite{Gingras_2014,Pore}.
 \\
One promising route toward uncovering fractionalization in frustrated magnets is the experimental identification of quantum spin liquid (QSL) states~\cite{ANDERSON1973153,Lancaster03042023}. In these highly frustrated systems, electron spins remain highly entangled and disordered even at zero temperature, leading to fractionalized excitations such as spinons and Majorana
fermions~\cite{RevModPhys.89.025003,Kitaev_2006}. These emergent quasiparticles not only deepen our understanding of many-body quantum physics but also hold promise for transformative applications, including quantum computing~\cite{Tokura2017,RevModPhys.80.1083}.\\
The search for QSLs in real magnetic materials was initiated by the pioneering proposal of resonating valence bond (RVB) states by P.~W.~Anderson and P.~Fazekas~\cite{ANDERSON1973153,Fazekas1974-FAZOTG}. Their seminal  work suggests that in geometrically frustrated systems such as triangular and kagome lattices,  strong spin frustration can destabilize long-range magnetic order, while promoting local valence-bond singlets~\cite{Balents2010,KHATUA20231}. These singlet pairs can dynamically rearrange their partners, ultimately giving rise to a liquid-like quantum state, termed a QSL state, without any symmetry breaking even as $T\rightarrow 0$~\cite{RevModPhys.89.025003}. \\ \\ When a singlet pair or a valence bond, breaks apart, it creates two unpaired spins~\cite{Savary_2016,Kim2006}. These unpaired spins form low-energy excitations known as spinons, which carry only half the spin of an electron but no electric charge. The separated charge degrees of freedom are known as holons, which carry the electron’s charge without its spin~\cite{RevModPhys.78.17,PhysRevB.83.224508}. In QSL, spinons can propagate throughout the lattice by rearranging nearby valence bonds, allowing them to move freely through the system~\cite{scienceaay0668,annurev401}.  Frustrated triangular and kagome lattices have been extensively studied as ideal platforms for realizing QSL states~\cite{RevModPhys.89.025003,Li_2020}. However, identifying definitive experimental signatures remains challenging in real materials and often falls short of theoretical expectations.\\
\begin{figure*}
	\centering
	\includegraphics[width=\linewidth]{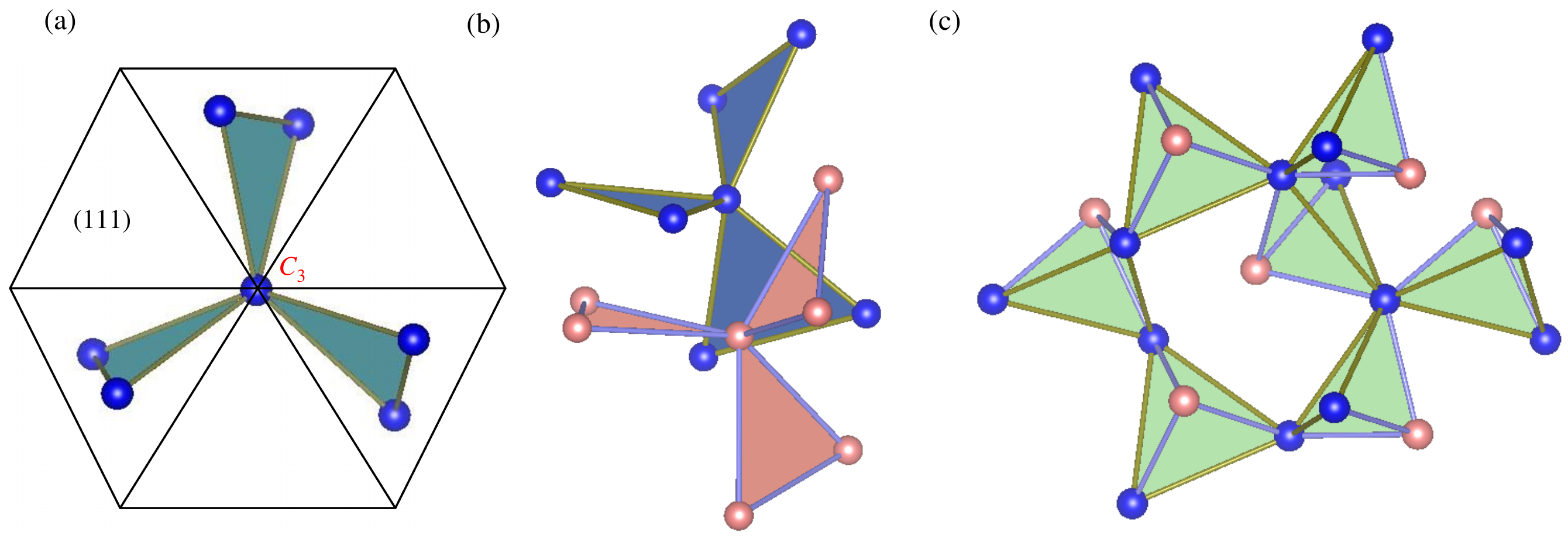}
	\caption{ (a) Schematic projection of the chiral trillium lattice (space group $P$2$_1$3), consisting of a three-dimensional network of corner-sharing equilateral triangles, onto the (111) plane, viewed along the $C_3$ rotational axis.   (b) Schematic of an interpenetrating double trillium lattice.
		(c) Geometry of a hypertrillium lattice, arising from the specific bond lengths between magnetic ions in the double trillium lattice.    }{\label{st}}.
\end{figure*}
In three dimensions (3D), pyrochlore lattices extend frustration-driven physics by incorporating strong Ising anisotropy along the local $\langle 111 \rangle$ directions~\cite{science4761,PhysRevLett.83.1854}.  A notable example is  3D  classical spin liquid (CSL) or classical spin ice  state in these lattices, where the magnetic moments mimic the proton disorder in water ice~\cite{doi:10.1021/ja01315a102}. This analogy gives rise to unusual low-temperature phenomena in real pyrochlore magnets, such as Ho$_{2}$Ti$_{2}$O$_{7}$ and Dy$_{2}$Ti$_{2}$O$_{7}$, including the emergence of magnetic monopole excitations and the existence of zero-point entropy~\cite{https168,Bramwell2020}. The latter refers to a vast number of degenerate ground states that remain even at absolute zero temperature~\cite{Ramirez1999}.\\
In spin ice, spins on tetrahedral units obey the ``two-in, two-out" ice rule, forming a disordered yet highly constrained magnetic configuration~\cite{science4761}. The creation of a ``three-in, one-out" state through spin flip generates monopole-like excitations that behave as isolated magnetic charges~\cite{Castelnovo2008,annurev070909-104138}. These excitations can propagate through the lattice and have been directly probed using neutron scattering~\cite{annurev}. However, the limited availability of suitable 3D pyrochlore materials remains a significant barrier to the further exploration of monopole physics and zero-point entropy  ~\cite{Pomaranski2013}.\\
More recently, attention has turned to kinetic frustration in the dilute-doping regime of half-filled Mott insulators as a route to realizing the RVB state in 3D frustrated lattices~\cite{Glittum2025}. Meanwhile, the search for QSL candidates has also revealed a range of other quantum phases and transitions. These include the realization of field-induced 1/9th magnetization plateaus~\cite{PhysRevLett.132.226701,Jeon2024}, spin nematic phases~\cite{Fogh2024}, Bose–Einstein condensation of two-magnon bound states~\cite{Sheng2025}, Kondo screening in QSL candidates~\cite{Lee2023,Gomilek2019}, supersolid phases~\cite{Xiang2024}, and topologically non-trivial spin textures such as skyrmions~\cite{Kurumaji914}.\\
In this context, expanding research beyond traditional frustrated magnets based on triangular, kagome, and pyrochlore lattices is essential for exploring a broader range of frustration-driven quantum states.  Notably, the trillium lattice, a chiral network composed of three equilateral triangles, offers a distinctive platform for realizing magnetic states rarely observed in conventional frustrated systems, as it combines geometric frustration with intrinsic structural chirality~\cite{Simonet2012}.

\subsection{Nexus between geometric frustration and lattice chirality}
\textcolor{black}{Intertwining frustration with chirality can create an environment for complex quantum and classical chiral phases distinct from both conventional chiral magnets and non-chiral frustrated magnets \cite{PhysRevLett.130.136701,Yokouchi2017}.
In non-centrosymmetric lattices, spin fluctuations arising from the frustrated triangular motifs can acquire a chiral character due to noncollinear spin configurations~\cite{Simonet2012,Bordcs2012,Kim2024}. These fluctuations are characterized by the scalar spin chirality,
	$\langle \mathbf{S}_{i} \cdot (\mathbf{S}_{j} \times \mathbf{S}_{k}) \rangle$,
	where $\mathbf{S}_{i}$, $\mathbf{S}_{j}$, and $\mathbf{S}_{k}$ denote localized spins on neighboring sites~\cite{PhysRevLett.130.136701}. \\
	Such chiral fluctuations have been observed to drive a large anomalous Hall effect in the triangular-lattice antiferromagnet PdCrO$_{2}$ \cite{Jeon2024}, while in the chiral trillium lattice MnSi they manifest as prominent nonreciprocal and nonlinear responses, such as the electrical magnetochiral effect~\cite{Yokouchi2017}. In metallic systems like Nd$_{2}$Mo$_{2}$O$_{7}$, electron hopping between noncollinear spins with finite chirality generates a Berry phase \cite{sciencechiral}, which acts as a fictitious magnetic field and induces anomalous charge dynamics. In localized magnets, spin chirality is likewise expected to influence spin and thermal excitations, detectable via thermal Hall measurements \cite{PhysRevB.103.224419,PhysRevLett.104.066403,PhysRevX.8.031032}, and may even stabilize chiral spin-liquid states in Mott insulators \cite{PhysRevB.111.205115}. Such quantum and classical spin liquids are marked by the absence of long-range magnetic order while breaking time-reversal and parity symmetries~\cite{Lozano2024,PhysRevLett.119.127204}.\\
Since chiral magnets are defined by their lack of inversion symmetry, they often entail Dzyaloshinskii–Moriya interactions (DMIs) that favor noncoplanar, helical, skyrmion, or topological chiral spin states~\cite{PhysRevLett.130.136701,Tokura2018}.
Notably, recent theory predicts that skyrmions and a chiral spin liquid can coexist as distinct disordered chiral phases, whose interplay may give rise to an unconventional topological spin glass~\cite{PhysRevLett.130.106703}.
In this vein, the trillium lattice—through its complex spin interactions, non-centrosymmetric geometry, and structural chirality- can serve as a playground for discovering skyrmion phases~\cite{science1166767}, spin-ice states~\cite{PhysRevB.82.014410}, chiral magnetic order~\cite{Cheong2022}, chiral spin liquids~\cite{Lozano2024}, and even chiral phonons.}
 \\

 In this review, we introduce the existing library of materials based on the trillium lattice, present theoretical insights from model calculations, and examine experimental findings observed in real compounds. Finally, we explore the prospects for uncovering new physics enabled by this unique and highly frustrated lattice platform.
\section{Trillium materials}
The trillium lattice structure was initially discovered in intermetallic transition-metal silicides, including MnSi, FeSi, and CoSi, all of which adopt the cubic B20 crystal structure (space group $P2_{1}$3)~\cite{science1166767,bradley2009mathematical,PhysRevB.74.224441,pnas1806910115}. This space group features a three-fold rotation axis and two special screw rotations that combine rotation and translation. A notable feature of these B20-type compounds is the presence of two interpenetrating trillium lattices. Each trillium lattice forms a chiral network of interconnected equilateral triangles (see figure~\ref{st}(a) and (b)), resembling a Canadian trillium flower and notably lacking inversion symmetry.\\
In such chiral magnets, the absence of inversion symmetry leads to the DMI, described by the antisymmetric exchange Hamiltonian $\mathcal{H}$ = $\sum_{\langle\langle i,j \rangle\rangle} \mathbf{D}_{ij} \cdot (\mathbf{S}_i \times \mathbf{S}_j)$ \cite{DZYALOSHINSKY1958241,PhysRev.120.91,Ham2021}. Here, the DM vector $\mathbf{D}_{ij}$ is antisymmetric under the exchange of sites $i$ and $j$, and its orientation is determined  by the crystal symmetry. Consequently, B20 materials commonly exhibit  non-collinear magnetic phases, such as helical spin structures~\cite{Bode2007} and magnetic skyrmions~\cite{ler2006}.
For instance, MnSi develops DMI-induced helical magnetic order below its Curie temperature $T_{\rm C} \sim 29$~K. Under moderate magnetic fields and near $T_{\rm C}$, it stabilizes topologically protected skyrmions~\cite{ISHIKAWA1976525}. Another trillium lattice is EuPtSi (Eu$^{2+}$;$S = 7/2$), which exhibits strong magnetic fluctuations above 4~K, its antiferromagnetic ordering temperature and hosts a field-induced skyrmion phase~\cite{Kakihana2017,PhysRevB.104.045145}. It is worth noting that, although two trillium networks exist in these intermetallic chiral magnets, magnetism is governed by a single trillium lattice formed by the magnetic ions, while the other lattice consists of non-magnetic atoms.
\\ \\
 \begin{table}
	\renewcommand{\arraystretch}{1.2}
	\caption{List of promising trillium lattice compounds that crystallize in the cubic B20 structure (space group $P$2$_1$3), identified as potential candidates for future investigations.}
	\label{jk}
	\centering
	\begin{tabular}{| m{1.25cm} | m{6.1cm} |}
		\hline
		\textbf{Spin} & \textbf{Compound and Reference} \\
		\hline
		$S = 1/2$ &
		\begin{tabular}[t]{@{}l@{}}
			K$_{2}$Ti$^{3.5+}_{2}$(PO$_{4}$)$_{3}$ \cite{https00059} \\
			(NH$_{4}$)(H$_{3}$O)Ti$^{3+}$Ti$^{4+}$(PO$_{4}$)$_{3}$ \cite{Fu:hb6237}
		\end{tabular} \\
		\hline
		$S = 1$ &
		\begin{tabular}[t]{@{}l@{}}
			Cs$_{2}$Ni$_{2}$(MoO$_{4}$)$_{3}$ \cite{Zolotova2011},	Ba$_{3}$V$^{3+}_{4}$(PO$_{4}$)$_{3}$ \cite{Dross:br6140} \\
			BaV$^{3+}$Ti(PO$_{4}$)$_{3}$,	Ba$_{1.5}$V$^{3+}$(PO$_{4}$)$_{3}$ \cite{Orlova2011} \\
			Pb$_{1.5}$V$^{3+}$(PO$_{4}$)$_{3}$ \cite{SHPANCHENKO20051569} \\
			
		\end{tabular} \\
		\hline
		$S = 3/2$ &
		K$_{2}$Co$^{2+}_{2}$(BeF$_{4}$)$_{3}$ \cite{LEFUR1969601} \\
		\hline
		$S = 2$ &
		K$_{2}$Fe$^{2+}_{2}$(SO$_{4}$)$_{3}$ \cite{DRISCOLL2020121363} \\
		\hline
		$S = 5/2$ &
		\begin{tabular}[t]{@{}l@{}}
			K$_{2}$Mn$^{2+}_{2}$(SO$_{4}$)$_{3}$,	Rb$_{2}$Mn$^{2+}_{2}$(MoO$_{4}$)$_{3}$ \cite{Bouzidi:ru2059} \\
			Pb$_{1.5}$Fe$^{3+}_{2}$(PO$_{4}$)$_{3}$ \cite{BOYA2025173302} \\
			K$_{2}$Fe$^{3+}_{2}$(MoO$_{4}$)(PO$_{4}$)$_{2}$ \cite{Slobodyanik2012}
		\end{tabular} \\
		\hline
		$J_{\rm eff} = \frac{1}{2}$ &
		\begin{tabular}[t]{@{}l@{}}
			K$_{2}$Yb$^{3+}$Ti(PO$_{4}$)$_{3}$ \cite{Norberg:os0095} \\
			KBa$R^{3+}_{2}$(PO$_{4}$)$_{3}$ ($R$ = Yb, Dy, Ho) \cite{Orlova2011} \\
		\end{tabular} \\
		\hline
		$J = 4$ &
		K$_{2}$PrZr(PO$_{4}$)$_{3}$ \cite{Orlova2011} \\
		\hline
		$J = 6$ &
		KBaTm$^{3+}_{2}$(PO$_{4}$)$_{3}$ \cite{Orlova2011} \\
		\hline
		$J = \frac{15}{2}$ &
		KBaEr$^{3+}_{2}$(PO$_{4}$)$_{3}$ \cite{Orlova2011} \\
		\hline
	\end{tabular}
\end{table}
  \begin{figure*}[t]
	\centering
	\includegraphics[width=\linewidth]{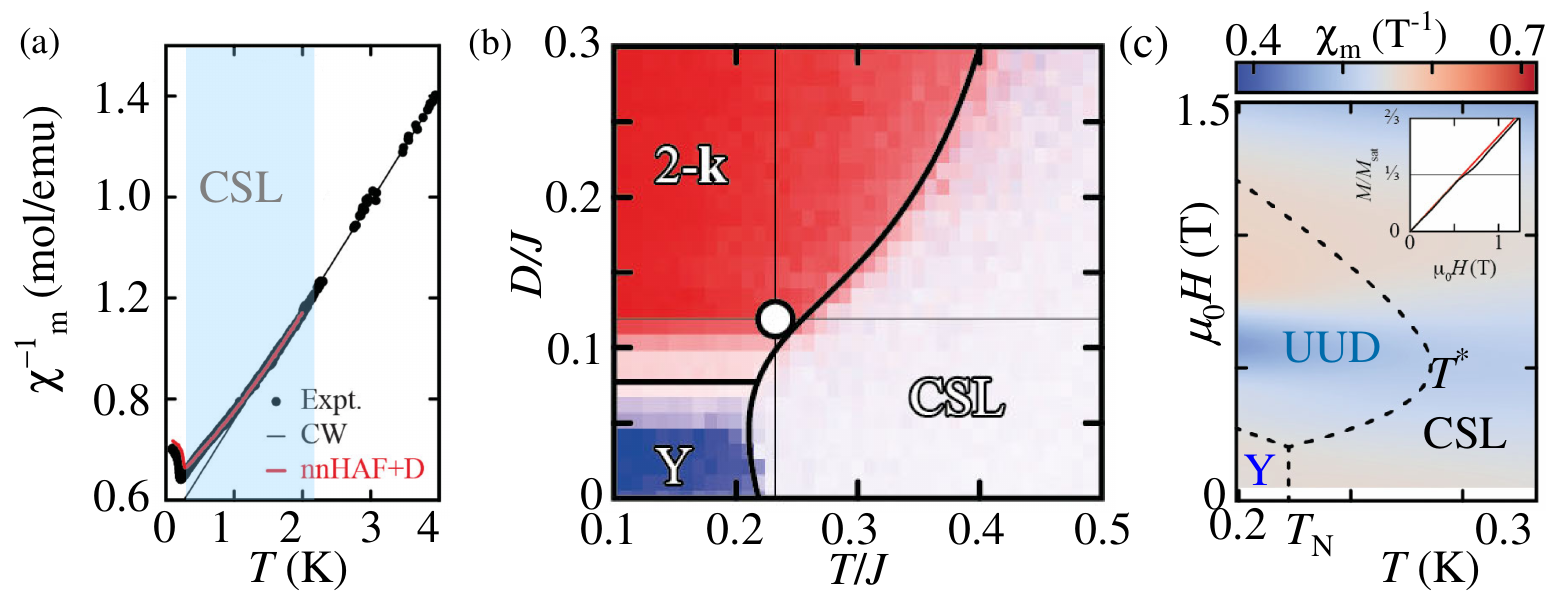}
	\caption{(a) Temperature dependence of the inverse magnetic susceptibility of Na[Mn(HCOO)$_3$] at 100\,Oe with a Curie-Weiss (CW) fit. The pale sky-blue shaded region indicates the classical spin liquid (CSL) regime, where the inverse magnetic susceptibility (red line) is obtained from classical Monte Carlo calculations, as discussed in the text. (b) Dipolar-to-Heisenberg interaction ratio ($D/J$) versus reduced temperature ($T/J$) phase diagram of a trillium lattice, with phase boundaries determined by classical
		Monte Carlo simulations. The intensity of red and blue shading reflects the strength of magnetic scattering associated with the 2-K and Y phases, respectively, while the empty circle indicates the position of Na[Mn(HCOO)$_3$] (c) Magnetic field--temperature phase diagram with a background contour map of $\chi_{m} = d(M/M_{\mathrm{sat}})/d(\mu_{0}H)$. The inset shows the isothermal magnetization at 100\,mK, with experimental data in black and a low-field linear fit in red. The field-induced UUD ordered phase, the 120$^{\circ}$ Y phase, and the CSL phase are separated by a dotted curve that closely follows the expected phase boundary from the antiferromagnetic Heisenberg model with an additional Zeeman term. Adapted from \cite{PhysRevLett.128.177201} with permission APS.
		
	}{\label{CSLMn}}.
\end{figure*}
 Beyond the B20-type metallic framework, a recent surge of interest in oxide-based frustrated systems has opened up promising new directions for engineering trillium lattices. Notably, this approach enables the exploration of novel magnetic phases governed by dominant Heisenberg exchange interactions with subdominant DM interactions.  A progress in this direction is the discovery of the first trillium lattice system governed primarily by Heisenberg interactions: sodium manganese(II) formate, $\mathrm{Na[Mn(HCOO)_3]}$, where magnetic $\mathrm{Mn^{2+}}$ ($S = \frac{5}{2}$) ions form a trillium lattice within the chiral cubic space group $P2_1 3$~\cite{PhysRevLett.128.177201}.
 \\
Another prominent class of trillium lattice compounds is found in the langbeinite family, which crystallizes in the non-centrosymmetric cubic space group $P2_1 3$, with the general formula $M_2M'_2(\mathrm{SO}_4)_3$ ($M^{+}$ = K, Cs; $M'^{2+}$ = Ni, Co, Mn). In these compounds, the magnetic $M'^{2+}$ ions occupy two crystallographically distinct sites, leading to a double trillium lattice without anti-site disorder between constituent atoms~\cite{PhysRevLett.127.157204,Zemann:a02028}(figure~\ref{st}(b)). Furthermore, when two independent trillium lattices are connected through specific exchange pathways, they form a hypertrillium spin topology composed of a 3D network of corner-sharing tetrahedra, as illustrated in figure~\ref{st}(c).
 This unique arrangement provides a structurally rich and chemically versatile platform for exploring frustration-driven magnetic phenomena in strongly correlated magnets.\\
Beyond sulfates, phosphate-based analogs such as $ABM''_{2}(\mathrm{PO}_4)_3$ ($A^{+}$ = Cs, Rb, Na, K; $B^{2+}$ = Sr, Ba; $M''^{3+}$ = Fe, Cr, Er)~\cite{HIDOURI2012145,BATTLE198616} and $A_{2}M''N''(\mathrm{PO}_4)_3$ ($M''$ = Fe, Cr, Yb, Er; $N''^{4+}$ = Ti, Sn) further expand this family, offering greater chemical flexibility and tunability of magnetic interactions~\cite{Orlova2011,Norberg:os0095,Boudjada:a15338}. Similar to the arrangement of divalent magnetic ions forming a double trillium lattice in the sulfate langbeinites, trivalent magnetic ions form a double trillium lattice in the phosphate analogues. In the $ABM''_{2}(\mathrm{PO}_4)_3$ series, the magnetic $M''$  site is fully ordered with no site dilution, although partial site mixing may occur between the non-magnetic $A$ and $B$ positions~\cite{PhysRevB.110.224405}. In contrast, the $A_{2}M''N''(\mathrm{PO}_4)_3$ compounds exhibit site mixing between magnetic $M''^{3+}$ and non-magnetic $N''^{4+}$ ions. This offers a unique advantage: by choosing compositions where one site is predominantly magnetic and the other largely non-magnetic (or vice versa), the spin topology can be tuned from a double trillium lattice to a nearly single trillium framework in the absence of random occupation. In addition, several phosphate-based langbeinite compounds have been identified and thoroughly discussed by Orlova \textit{et al}. in Refs.~\cite{Orlova2011,BOYA2025173302,Kub2024}. A list of unexplored but potentially promising chiral trillium lattice candidates is summarized in Table~\ref{jk}, highlighting materials that warrant future experimental and theoretical investigation. Notably, the spin number in these systems is highly tunable, ranging from $S$ = 1/2 to $S$ = 5/2, as well as effective total angular momentum values $J_{\rm eff}$ = 1/2, 4, 6, and 15/2.
 \\
  \begin{table*}[t]
 	\centering
 	\caption{Summary of theoretical approaches and quantum phenomena proposed for the trillium lattice, where some results are limited by \textcolor{black}{finite-size effects}, along with key open questions for real materials.}
 	\label{theorysummary}
 	\renewcommand{\arraystretch}{1.3}
 	\begin{tabular}{|>{\centering\arraybackslash}p{2.3cm}|>{\centering\arraybackslash}p{2.5 cm}|>{\centering\arraybackslash}p{4.5cm}|>{\centering\arraybackslash}p{6.0cm}|}
 		\hline
 		\textbf{\centering Approach} & \textbf{Key Findings} & \textbf{Proposed Phenomena} & \textbf{Open questions for real materials} \\
 		\hline
 		Heisenberg model & $120^\circ$ helical order & Degenerate ground state, cooperative paramagnetism~\cite{PhysRevB.74.224441} & \textcolor{black}{Role of quantum fluctuations, spin–orbit coupling, and chiral interactions} \\
 		\hline
 		Heisenberg + dipolar model & Y-phase and 2-$\mathbf{k}$ phase & Classical spin liquid, 1/3 magnetization plateau~\cite{PhysRevLett.128.177201} & \textcolor{black}{Effects of single-ion anisotropy, finite DM interactions, and lattice defects} \\
 		\hline
 		Ferromagnetic Ising model & Residual entropy $S/N = \ln(3/2)$ & Classical spin-ice state, monopole-like excitations~\cite{PhysRevB.82.014410} & \textcolor{black}{Influence of transverse fluctuations and dipolar interactions} \\
 		\hline
 		Newman–Moore model & Fractal spin liquid & Fractal symmetry and fracton excitations~\cite{PhysRevB.105.224410} & \textcolor{black}{How excitations couple to probes (neutron, $\mu$SR, thermal Hall)} \\
 		\hline
 		Projective symmetry group (PSG) & $\mathbb{Z}_2$ and U(1) spin liquids & Quantum spin liquid, spinon Fermi surface~\cite{3nmp1vt2} & \textcolor{black}{Stability beyond mean-field; effect of perturbations (e.g., DMI, exchange anisotropy)} \\
 		\hline
 	\end{tabular}
 \end{table*}
\section{Theoretical approaches}
In frustrated magnets, competing interactions often generate a highly degenerate classical ground state \cite{Starykh}, which can be lifted by thermal or quantum fluctuations through the order-by-disorder mechanism~\cite{villain}. The structure and dimensionality of the ground-state manifold determine the nature of these fluctuations: certain configurations allow softer modes or a higher density of low-energy excitations, which are entropically or quantum mechanically favored. Consequently, fluctuations stabilize specific ordered states within the degenerate manifold by minimizing either the free energy (thermal case) or the zero-point energy (quantum case)~\cite{PhysRevB.111.184434,PhysRevLett.62.2056}. Thus, the geometry of classical ground states governs fluctuation effects and enables emergent order in frustrated systems~\cite{PhysRevB.102.214424,PhysRevLett.122.017201}.\\ \\
Unlike the well-studied pyrochlore and kagome lattices, the trillium lattice has received relatively little theoretical attention regarding its ground state. Classical Heisenberg model studies with nearest-neighbor antiferromagnetic interactions predict a $120^\circ$ helical order with wave vector $\mathbf{k} = [1/3,,0,,0]$, and thermal fluctuations drive a phase transition at $T_{\rm N} \approx 0.21J$, where $J$ is the interaction strength \cite{PhysRevB.74.224441}. Yet, conflicting theoretical results leave the precise ground state unresolved. Moreover, it has been suggested that, akin to kagome and pyrochlore antiferromagnets, the trillium lattice may host a CSL state between $T_{\rm N}$ and the Curie-Weiss temperature $|\theta_{\rm CW}|$, where spins form a cooperative paramagnetic state with short-range correlations~\cite{PhysRevB.78.014404}.
\\
Indeed, such a CSL regime has been proposed in the trillium lattice Na[Mn(HCOO)$_3$] (Mn$^{2+}$, $S=5/2$), which shows dominant antiferromagnetic interactions with $\theta_{\rm CW} = -2.3$\,K and a N\'eel temperature $T_{\rm N} = 0.22$\,K (see figure~\ref{CSLMn}(a)) \cite{PhysRevLett.128.177201}. The deviation of the inverse magnetic susceptibility from CW behavior in the temperature range $T_{\rm N} < T < \theta_{\rm CW}$ is interpreted as the presence of a CSL state (blue shaded region in figure~\ref{CSLMn}(a)). Further evidence for the CSL state was provided by strong diffuse scattering observed  at $T$ = 1.5 K ($>T_{\rm N}$) in  polarized neutron scattering experiments~\cite{PhysRevLett.128.177201}.\\ In order to further examine   the magnetic structure below $T_{\rm N}$, neutron diffraction measurements were carried out by   Bulled \textit{et al.} at  0.1 K ($T < T_{\rm N}$). Their results revealed a magnetic structure inconsistent with the previously proposed simple 120$^\circ$ (Y) phase. Instead, they identified a 2-$\mathbf{k}$ phase with wave vector $\mathbf{k}$ = [1/2, 0, 0], implying the presence of  additional interactions. To address this, Bulled \textit{et al.}~\cite{PhysRevLett.128.177201} modeled the magnetic susceptibility data using the spin Hamiltonian \[
\mathcal{H}
= J \sum_{\langle i , j \rangle} \mathbf{S}_i \cdot \mathbf{S}_j
+ D \sum_{i > j} \frac{\mathbf{S}_i \cdot \mathbf{S}_j + 3(\mathbf{S}_i \cdot \hat{\mathbf{r}}_{ij})(\mathbf{S}_j \cdot \hat{\mathbf{r}}_{ij})}{\left(r_{ij} / r_1\right)^3},
\] where $J \sim 1$~K is the nearest-neighbor exchange strength and $D = 0.118$~K represents the dipolar interactions. As is  evident from the phase diagram
of $D/J$ vs $T/J$ in figure~\ref{CSLMn}(b), three distinct phases were identified: the Y-phase for small $D/J$, a 2-$\mathbf{k}$ phase  for $D/J \geq 0.08$, and a CSL regime at higher $T/J$. The position corresponding to the Na[Mn(HCOO)$_3$] compound is marked by a white circle, with vertical and horizontal lines denoting its specific values of $D/J$ and $T/J$ (figure~\ref{CSLMn}(b)).\\
In 2D triangular and kagome lattices, magnetization plateaus arise as quantum and thermal fluctuations stabilize particular spin configurations~\cite{Nishimoto2013,Kamiya2018}. A well-known case is the 1/3 “up-up-down” (UUD) state, where one-third of the spins align with the external field while the rest remain frustrated. In the 3D trillium lattice compound Na[Mn(HCOO)$_3$], a $1/3$ magnetization pseudoplateau appears at $\mu_0 H \approx 0.6$~T, evidenced by a dip in differential susceptibility [figure~\ref{CSLMn}(c)] \cite{PhysRevLett.128.177201}. This mirrors the UUD state in 2D frustrated systems and is supported by Monte Carlo simulations. The plateau persists up to $T \approx 0.29$~K, consistent with theoretical predictions [figure~\ref{CSLMn}(c)]~\cite{PhysRevLett.128.177201}.
\\
Having discussed the antiferromagnetic Heisenberg model with dipolar interactions on the trillium lattice, we now consider the ferromagnetic regime and its potential to host spin-ice physics~\cite{Bramwell1495}. In 3D pyrochlore lattices, ferromagnetic Ising spins along the local $\langle 111 \rangle$ axes obey the “two-in/two-out” ice rule, while in 2D kagome lattices a two-in/one-out configuration has been proposed, as in HoAgGe~\cite{sciencegeg}. Whether the trillium lattice supports a similar state remains under debate. Classical Monte Carlo simulations of the ferromagnetic Ising model on the trillium lattice reveal a degenerate ground state consistent with the two-in/one-out rule and a residual entropy $S/N = \ln(3/2)$, matching Pauling’s estimate~\cite{doi:10.1021/ja01315a102,PhysRevB.82.014410}. Although no experimental realization exists yet, the trillium lattice composed of rare-earth ions is a promising platform for monopole-like excitations. \textcolor{black}{Notably, the positive CW temperature in EuPtSi suggests proximity to a ferromagnetic Ising model and motivates its proposal as a spin-ice candidate~\cite{PhysRevB.82.014410}, though experimental confirmation is limited and the origin of frustrated magnetism above $T_\mathrm{N} = 4$ K remains to be clarified (Section~\ref{SKYR})}.\\

Apart from bilinear spin interactions, a classical Ising model with three-spin interactions on the trillium lattice—the Newman-Moore model—has been studied~\cite{PhysRevE.60.5068}. Here, spins on each triangular plaquette interact via $\mathcal{H} = - \sum_{\Delta i,j,k} \sigma_i \sigma_j \sigma_k$, with $\sigma_i = \pm 1$~\cite{PhysRevB.105.224410}. The ground states and low-energy excitations exhibit fractal properties due to exact fractal symmetries acting on subsets of spins, resulting in a classical fractal spin liquid. Excitations, or fractons, are localized defects created by flipping fractal spin subsets~\cite{annurev04,PhysRevB.111.134413}. These fractons are immobile and separated by energy barriers growing logarithmically with system size, leading to slow, glassy dynamics at low temperatures~\cite{PhysRevB.105.224410}. \textcolor{black}{While realizing the Newman-Moore model in bulk materials is challenging, quantum simulators may offer a feasible platform.}\\

\begin{figure}[t]
	\centering
	\includegraphics[width=\linewidth]{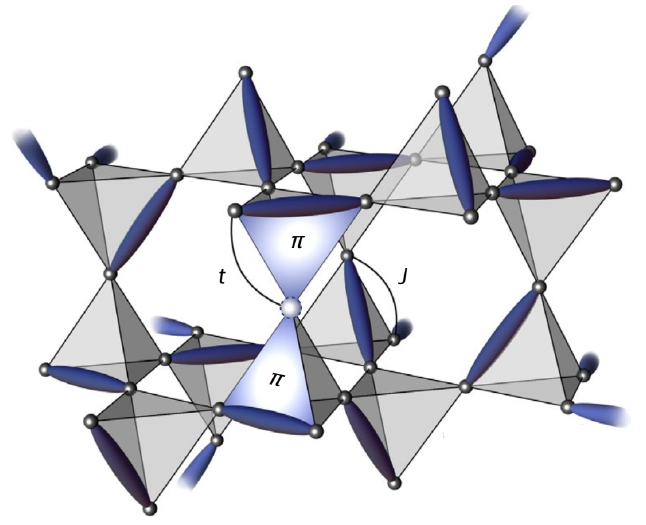}
	\caption{(a) Schematic of dimer-singlet coverings (shown as dark thin ellipsoids) on the pyrochlore lattice, with one singlet per tetrahedron. It also shows a hole (light-colored sphere) moving through the lattice by binding two $\pi$-fluxes to it.  Adapted from \cite{Glittum2025} with permission NPG.  }{\label{RVB3d}}.
\end{figure}
Moving beyond classical models, strong frustration and quantum fluctuations in 3D lattices can suppress symmetry breaking and stabilize topologically ordered phases such as QSLs, which host emergent fractional excitations like spinons. While fractionalization is well established in 2D, its realization in 3D remains debated. The RVB concept, central to QSLs, has recently been extended to 3D pyrochlores, where kinetic—rather than exchange—frustration may stabilize an RVB state~\cite{Glittum2025}.\\

The concept of kinetic frustration originates from Nagaoka’s theorem, which states that in bipartite lattices with infinite on-site Coulomb repulsion, adding a single electron or hole aligns all spins (ferromagnetism) to allow dopant motion and lower its kinetic energy~\cite{PhysRev.147.392,PhysRevLett.86.3396}. On non-bipartite lattices, dopant motion interferes with itself, preventing kinetic energy gain even in a fully aligned state. This quantum phenomenon, kinetic frustration, suppresses full spin alignment and can stabilize unconventional antiferromagnetic states, such as Haerter-Shastry-type order~\cite{PhysRevLett.95.087202,PhysRevResearch.5.L022048,PhysRevLett.112.187204}.\\

Recently, Glittum \textit{et al.}\cite{Glittum2025} extended kinetic frustration to 3D pyrochlores, showing that at half-filling and in the large-$U$, $J=0$ limit, the lattice can host an RVB state—a quantum superposition of dimer coverings with one spin singlet per tetrahedron (figure~\ref{RVB3d}) described by the $t$-$J$ model,
$\mathcal{H} = -t \sum_{\langle i,j \rangle, \sigma} \left[ \hat{c}^{\dagger}_{i\sigma} \hat{c}_{j\sigma} + \mathrm{h.c.}
 \right] + J \sum_{\langle i , j \rangle} \mathbf{S}_i \cdot \mathbf{S}_j ,$
where $\hat{c}^{\dagger}{i\sigma}$ ($\hat{c}{i\sigma}$) forbid double occupancy. Upon hole doping, the system generates dynamic $\pi$-fluxes on triangular plaquettes, enabling hole delocalization and lowering the kinetic energy to $-4t$, validating the 3D RVB picture~\cite{Glittum2025}.
Hole motion breaks a singlet, producing two fractional excitations: a spinon (spin-only) and a holon (charge-only), with fluxes arising dynamically from dimer fluctuations (figure~\ref{RVB3d})~\cite{Glittum2025}. These results extend RVB physics from 2D to 3D lattices. While current theory focuses on single or double hole doping at $J = 0$, the multi-hole regime remains an open question. Moreover, similar RVB states may appear in the weak exchange limit of Mott insulators~\cite{Glittum2025}.\\

  \begin{figure*}
 	\centering
 	\includegraphics[width=\linewidth]{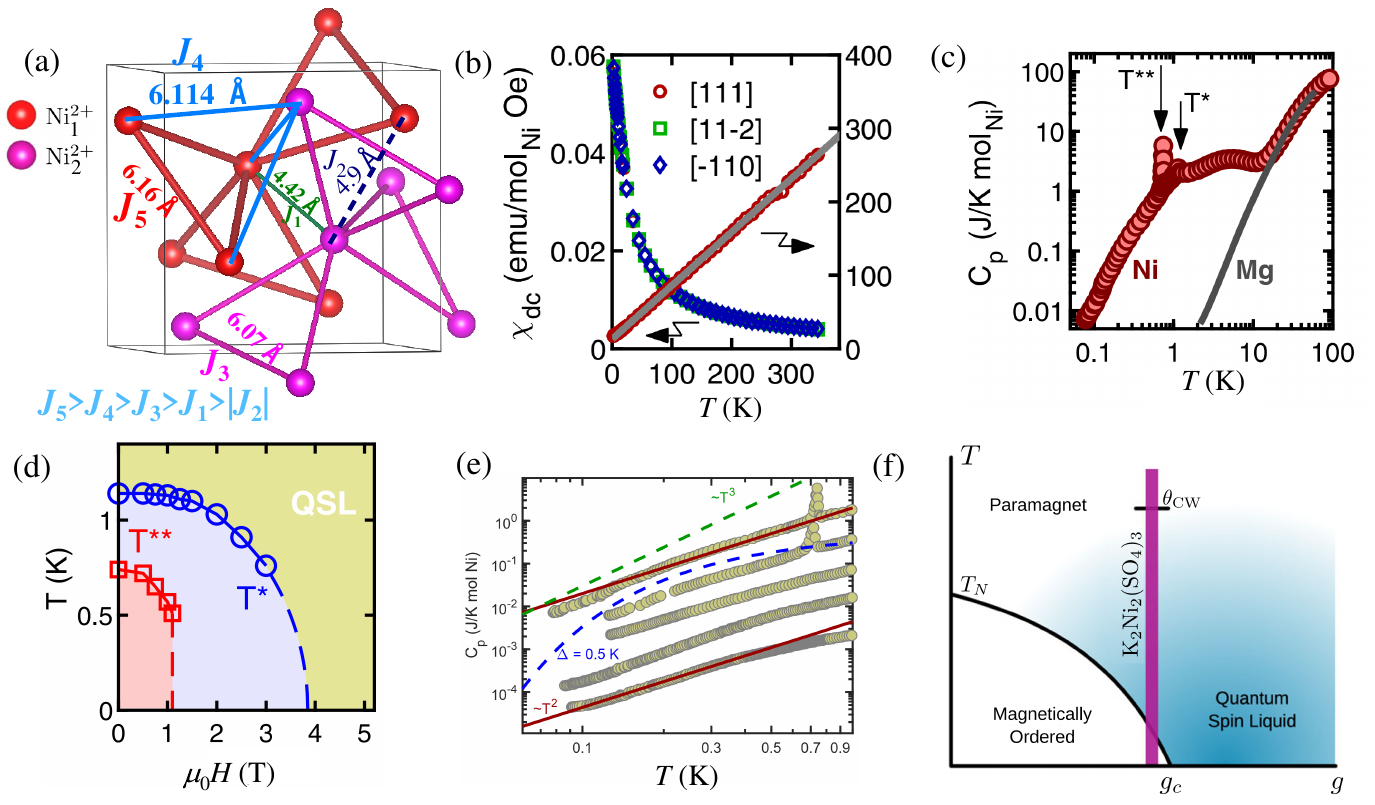}
 	\caption{(a) Schematic of the double trillium lattice formed by two sublattices of Ni$^{2+}$ ions in K$_{2}$Ni$_{2}$(PO$_{4}$)$_{3}$. Other atoms are omitted for clarity to highlight the two interpenetrating trillium networks. Among the five exchange interactions, only $J_4$ and $J_5$ dominate, jointly forming a hypertrillium lattice that induces strong frustration in the system.	(b) Temperature dependence of the magnetic susceptibility measured in an applied field of 0.1\,T along the three orthogonal directions of the cubic lattice (left $y$-axis), with the inverse magnetic susceptibility for $H$ $//$[111] plotted on the right $y$-axis.	(c) Temperature dependence of the specific heat, showing two anomalies at $T^{*} = 1.14$\,K and $T^{**} = 0.74$\,K. The solid line represents the specific heat of the nonmagnetic analogue K$_{2}$Mg$_{2}$(SO$_{4}$)$_{3}$. (d) Magnetic field--temperature phase diagram illustrating the suppression of the specific heat anomaly with increasing field.	(e) Temperature dependence of the specific heat at low temperatures under several magnetic fields: 0\,T, 0.5\,T, 1.5\,T, 7\,T, and 14\,T (top to bottom). The solid and dashed lines indicate the nature of the magnetic excitations as described in the text. 	(f) Schematic phase diagram showing temperature ($T$) versus tuning parameter ($g$), illustrating the proximity of K$_2$Ni$_2$(SO$_4$)$_3$ to a quantum critical point $g_c$. The material lies in a regime where quantum fluctuations suppress magnetic order ($T_{\rm N} \ll |\theta_{\rm CW}|$), leading to the emergence of a quantum spin liquid phase. The shaded region marks the quantum critical regime. Adapted from \cite{PhysRevLett.127.157204} and \cite{Gonzalez2024} with permission APS and NPG, respectively.
 		
 	}{\label{KNSO}}.
 \end{figure*}

 Beyond the pyrochlore lattice, the trillium lattice stands out as a complementary setting to explore RVB-related QSL physics in 3D. A major advance in the theoretical study of QSLs with topological order comes from the projective symmetry group (PSG) framework~\cite{PhysRevB.65.165113}, which classifies symmetric QSLs by incorporating lattice symmetries into low-energy effective theories—typically SU(2), U(1), or $\mathbb{Z}_2$ gauge structures~\cite{RevModPhys.78.17}. These gauge theories capture the long-range quantum entanglement intrinsic to QSL states. Recently, Li \textit{et al.}~\cite{3nmp1vt2} used the PSG approach to classify symmetric QSLs on the trillium lattice with its $P2_1 3$ space group symmetry. Their analysis reveals a limited set of QSLs: two gapless $\mathbb{Z}_2$ QSLs and one gapless U(1) QSL. The U(1) state features a spinon Fermi surface and lies near one of the $\mathbb{Z}_2$ phases in parameter space. All three proposed QSL states are gapless.
\\
\textcolor{black}{Theoretically, a rich variety of exotic quantum states have been predicted, as listed in Table~\ref{theorysummary}. Yet, experimental progress is limited by the lack of suitable materials. Real magnetic systems include subleading interactions such as DMI, chiral couplings, and single-ion anisotropy whose impact on the ground state remains unclear. Most models focus on a single trillium topology, whereas several langbeinite compounds realize coupled trillium lattices. Experimental studies on double-trillium systems, such as K$_2$Ni$_2$(SO$_4$)$_3$ \cite{PhysRevLett.127.157204} and KSrFe$_2$(PO$_4$)$_3$ \cite{10.1063/5.0096942}, show behaviors deviating from single-lattice expectations, highlighting the need for theoretical frameworks including inter-lattice couplings. By comparison, pyrochlore magnets exhibit well-known pinch-point features in neutron structure factors~\cite{science1177582}; analogous predictions for (double-)trillium lattices with chiral spin fluctuations would greatly advance understanding.\\
Disorder is another central issue. Nearly all known trillium compounds exhibit intrinsic disorder from site mixing or bond randomness, which can strongly affect their ground states. However, theoretical models incorporating such disorder remain scarce. Insights from 2D frustrated systems suggest that disorder can induce disorder-driven chiral order in kagome lattices~\cite{letouze2025chiralorderemergencedriven} or random-singlet states~\cite{PhysRevLett.127.127201,shimokaw}. Extending these ideas to 3D trillium and hypertrillium lattices represents a promising research direction. A key next step is to determine the percolation threshold of antisite disorder, which may distinguish clean-lattice behavior from disorder-dominated states~\cite{PhysRevB.106.075146}. 
In the pursuit of the trillium lattice, langbeinites have been recently identified as potential hypertrillium hosts, awaiting both theoretical and experimental investigation.}

\begin{table*}\label{Table}
		\caption{Summary of structural parameters, magnetic interactions, and exotic phenomena in selected trillium lattice compounds.}
	\centering
	\begin{tabular}{ | p{3.3cm} | p{3.8cm} | c | c | p{3cm} |>{\centering\arraybackslash} c | }
		\hline
		\parbox[c]{3.4cm}{\centering \textbf{Trillium compound}}  &
		\parbox[c]{4.2cm}{\centering \textbf{Lattice parameter,} \\ \textbf{Bond lengths}} &
		\textbf{$\theta_{\rm CW}$ (K)} &
		\textbf{$T_{\rm N}$ (K)} &
	\centering	\textbf{Phenomena} &
		\textbf{Ref.} \\  \hline
			\parbox[c]{4.1cm}{\centering MnSi \\ ($P$2$_1$3)}	  &
		\parbox[c]{4.2cm}{\centering $a = 4.558$~Å \\ Mn--Mn = 2.80~Å \\} &
		$\sim +30$ &
		$T_{\rm C}\sim 29.5$ &
	\centering{Skyrmion}  &
		\cite{PhysRevB.93.144419,PhysRevB.55.8330} \\ \hline
			\parbox[c]{4.1cm}{\centering FeSi] \\ ($P$2$_1$3)}	  &
		\parbox[c]{4.2cm}{\centering $a = 4.48$~Å \\ Fe--Fe =2.74~Å \\} &
		$> -400$ &
	-- &
		\centering{Chiral phonon}  &
		\cite{PhysRevB.38.6954,PhysRevLett.121.035302} \\ \hline
			\parbox[c]{4.1cm}{\centering EuPtSi] \\ ($P$2$_1$3)}	  &
		\parbox[c]{4.2cm}{\centering $a = 6.433$~Å \\ Eu--Eu = 3.94~Å \\} &
		$+4$ &
		4 &
		\centering{Skyrmion}  &
	\cite{Kakihana2017} \\ \hline
			\parbox[c]{4.1cm}{\centering Cu$_{2}$OSeO$_{3}$ \\ ($P$2$_1$3)}	  &
		\parbox[c]{4.2cm}{\centering $a = 8.91$~Å \\ Cu1--Cu1 = 5.48~Å \\ Cu2--Cu2 = 3.22~Å} &
		$+69$ &
		$T_{\rm C}$ $\sim$ 60 &
	\centering{Skyrmion}  &
		\cite{PhysRevB.78.094416,PhysRevLett.108.237204} \\ \hline
			\parbox[c]{4.1cm}{\centering Na[Mn(HCOO)$_{3}$] \\ ($P$2$_1$3)}	  &
		\parbox[c]{4.2cm}{\centering $a = 9.132$~Å \\ Mn--Mn = 5.6~Å \\} &
		$-2.3$ &
		0.223 &
	\centering{Classical spin liquid}  &
		\cite{PhysRevLett.128.177201} \\ \hline

		\parbox[c]{4.1cm}{\centering K$_2$Ni$_2$(SO$_4$)$_3$ \\ ($P$2$_1$3)}	  &
		\parbox[c]{4.2cm}{\centering $a = 9.818$~Å \\ Ni1--Ni1 = 6.12~Å \\ Ni2--Ni2 = 6.08~Å} &
		$-18$ &
		0.74, 1.14 &
		\centering{Quantum spin liquid}  &
		\cite{PhysRevLett.127.157204} \\ \hline
			\parbox[c]{4.2cm}{\centering K$_2$Co$_2$(SO$_4$)$_3$ \\ ($P$2$_1$3,$P2_{1}$)}	  &
		\parbox[c]{4.2cm}{\centering $a = 9.818$~Å \\ Co1--Co1 = 6.18~Å \\ Co2--Co2 = 6.137~Å} &
		$-39$ &
	$\sim$ 0.6 K &
	\centering{Multiferroicity} &
		\cite{cavatri} \\ \hline
			\parbox[c]{4.2cm}{\centering 	KSrFe$_{2}$(PO$_{4}$)$_{3}$ \\ ($P$2$_1$3)} &
		\parbox[c]{4.2cm}{\centering $a = 9.78$~Å\\ Fe1--Fe1 = 6.03~Å\\ Fe2--Fe2 = 6.10~Å} &
		$-70$ &
	-- &
	 \centering{Spin liquid} &
	\cite{10.1063/5.0096942} \\ \hline
	
		\parbox[c]{4.0cm}{\centering Cs$_{2}$Fe$_{2}$(MoO$_{4}$)$_{3}$ \\ ($P$2$_1$3)}	 &
	\parbox[c]{4.2cm}{\centering $a = 10.911$~Å\\ Fe1--Fe1 = 6.78~Å\\ Fe2--Fe2 = 6.69~Å} &
	$-22$ &
	$\sim$1 &
	\parbox[c]{3.4cm}{\centering Cooperative paramagnet \\ (1 K $\leq$ $T$ $\leq$ $|\theta_{\rm CW}|)$} &
	\cite{Kub2024} \\ \hline

		\parbox[c]{4.2cm}{\centering K$_2$FeSn(PO$_4$)$_3$ \\ ($P$2$_1$3)}	 &
		\parbox[c]{4.2cm}{\centering $a = 9.914$~Å\\ Fe1--Fe1 = 6.10~Å\\ Fe2--Fe2 = 6.17~Å} &
		$-43$ &
		2 &
		Spin fluctuations, local spin singlet &
		\cite{lmsf73hn} \\ \hline
			\parbox[c]{4.0cm}{\centering K$_2$CrTi(PO$_4$)$_3$ \\ ($P$2$_1$3)}	 &
		\parbox[c]{4.2cm}{\centering $a = 9.796$~Å\\ Cr1--Cr1 = 6.09~Å\\ Cr2--Cr2 = 6.02~Å} &
		$-23$ &
		4.3, 8 &
	\parbox[c]{3.4cm}{\centering Cooperative paramagnet \\ (8 K $\leq$ $T$ $\leq$ $|\theta_{\rm CW}|)$} &
		\cite{PhysRevB.109.184432} \\ \hline
			\parbox[c]{4.2cm}{\centering KBaCr$_{2}$(PO$_{4}$)$_{3}$ \\ ($P$2$_1$3)}	 &
		\parbox[c]{4.2cm}{\centering $a = 9.79$~Å\\ Cr1--Cr1 = 6.08~Å\\ Cr2--Cr2 = 6.026~Å} &
		$-23$ &
		7, 13.5 &
		\parbox[c]{3.1cm}{\centering Compensated   \\ antiferromagnetism} &
		\cite{PhysRevB.109.184432} \\ \hline
		
	\end{tabular}
	\label{table}
\end{table*}

\section{Experimental results}
 \subsection{{\normalfont\textbf{Field-induced QSL state in K$_{2}$Ni$_{2}$(SO$_{4}$)$_{3}$}}}

K$_{2}$Ni$_{2}$(SO$_{4}$)$_{3}$ is the first known member of the langbeinite family to exhibit intriguing spin dynamics caused by multiple competing magnetic interactions~\cite{PhysRevLett.127.157204,PhysRevLett.131.146701}. In the following, we provide a detailed account of the structure and magnetism of this system, as it serves as a representative prototype exhibiting features characteristic of the langbeinite family. K$_{2}$Ni$_{2}$(SO$_{4}$)$_{3}$ crystallizes in the cubic space group $P$2$_{1}$3 and contains two symmetrically inequivalent crystallographic sites for Ni$^{2+}$ ($S$ = 1) ions~(figure~\ref{KNSO}(a)). This compound is free from detectable atomic site mixing. Notably, the shortest Ni--Ni bond distance of 4.42\,\AA{} forms a dimer between Ni1 and Ni2 sites, while the individual Ni1 and Ni2 sublattices each form a trillium lattice with Ni--Ni bond distances of 6.16\,\AA{} and 6.07\,\AA{}, respectively~(figure~\ref{KNSO}(a)). Despite, the shorter Ni1–Ni2 dimer bond, exchange interactions within the trillium lattices are stronger.  The dominant magnetic exchange interactions are expected to occur via superexchange pathways of the form Ni--O--S--O--Ni. Based on the room-temperature structure and CW temperature (see below), density functional theory (DFT) calculations yield five exchange interactions, namely $J_1 = 0.42$ K, $J_2 = -0.16$ K, $J_3 = 1.09$ K, $J_4 = 5.38$ K, and $J_5 = 2.54$ K (see figure~\ref{KNSO}(a)).
 Among them, $J_4$ and $J_5$ are dominant and form a hypertrillium spin network, leading to enhanced spin frustration in K$_{2}$Ni$_{2}$(SO$_{4}$)$_{3}$~\cite{PhysRevLett.127.157204,Gonzalez2024}. A slight variation in these five exchange interactions has been reported by Yao \textit{et al}. though the strong $J_4$ and $J_5$ couplings remain robust~\cite{PhysRevLett.131.146701}.\\ Theoretical studies suggest that the ground state of a single trillium lattice should exhibit a variant of 120$^{\circ}$ magnetic order; however, this prediction awaits experimental validation as real candidate materials are subject to perturbation~\cite{PhysRevB.74.224441}. In contrast, the ground state of the interconnected double trillium lattice compound K$_{2}$Ni$_{2}$(SO$_{4}$)$_{3}$ is found to be more exotic than that proposed for the single trillium lattice. This could be due to the presence of several competing exchange interactions that may naturally give rise to strong quantum fluctuations, in a manner similar to that observed in low-dimensional materials~\cite{Vasiliev2018,KHATUA20231} \\ \\
Figure~\ref{KNSO}(b) shows the temperature dependent magnetic susceptibility of K$_{2}$Ni$_{2}$(SO$_{4}$)$_{3}$  under a $\mu_{0}H = 0.1$\,T field applied along three cubic directions (left $y$-axis of figure \ref{KNSO}(b)) \cite{PhysRevLett.127.157204}. No long-range magnetic order is found down to 2 K, showing no noticeable anisotropy. The CW analysis yields the CW temperature of $-18$ K, indicating antiferromagnetic interactions between Ni$^{2+}$ ($S$ = 1) moments (right $y$-axis of figure \ref{KNSO}(b)).
\\
Figure~\ref{KNSO}(c) shows the temperature dependence of the specific heat, alongside the phonon contribution (inferred from the non-magnetic sister compound), which dominates only above 20\,K. Below this temperature, magnetic contributions show a broad peak around 5\,K, a hallmark of short-range spin correlations. Such features are typical in low-dimensional magnets and are also observed in the 3D trillium lattice like QSL candidate PbCuTe$_2$O$_6$, which hosts several competing exchange interactions~\cite{PhysRevB.90.035141,PhysRevB.108.184415}, similar to K$_2$Ni$_2$(SO$_4$)$_3$~\cite{PhysRevLett.127.157204}. In contrast to PbCuTe$_2$O$_6$, K$_2$Ni$_2$(PO$_4$)$_3$ exhibits a narrow anomaly at $T^{*} = 1.14$\,K and a sharp $\lambda$-type feature at $T^{**} = 0.74$\,K. The asymmetric anomaly at $T^{**}$ is proposed to be a second-order phase transition associated with the low-temperature magnetic phase, while the symmetric anomaly at $T^{*}$, observed only in single crystals, is considered as a first-order phase transition.\\ \\
To figure out the magnetic ordering structure, Živković \textit{et al.} performed neutron powder diffraction measurements above and below $T^{**}$. At 0.1\,K, weak magnetic Bragg peaks for $Q < 1$\,\AA$^{-1}$ were observed~\cite{PhysRevLett.127.157204}, corresponding to three propagation vectors, indicating that magnetic frustration suppresses long-range order and permits multiple nearly degenerate structures. The application of a magnetic field gradually suppresses both anomalies, as shown in the temperature–field phase diagram (figure~\ref{KNSO}(d)), confirming their magnetic origin. The $T^{**}$ anomaly vanishes in fields $\mu_0 H \approx 1$\,T, while $T^{*}$ remains unaffected until $\mu_0 H \approx 4$\,T.
\\  \begin{figure*}[t]
 	\centering
 	\includegraphics[width=\linewidth]{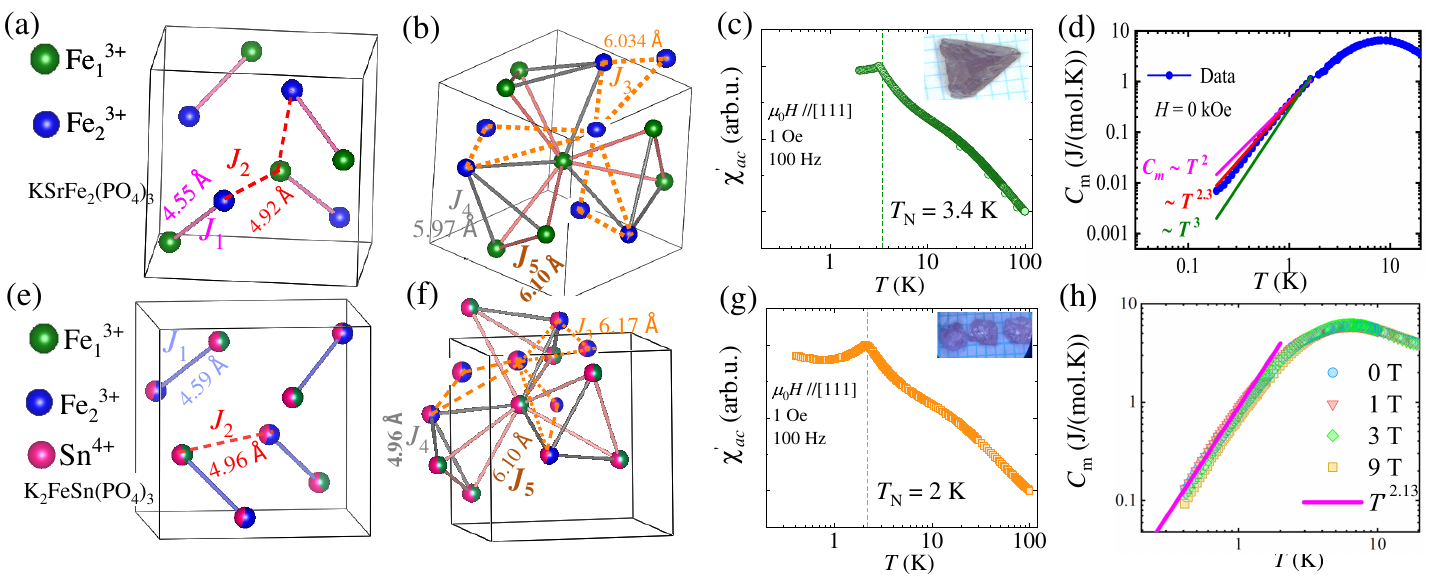}
 	\caption{The first row presents the structural and thermodynamic data of KSrFe$_{2}$(PO$_{4}$)$_{3}$, while the second row shows the corresponding data for K$_{2}$FeSn(PO$_{4}$)$_{3}$. (a) Schematic illustration of randomly oriented dimers formed by Fe1 and Fe2 sites, coupled via intra-dimer exchange interaction $J_1$, which are further linked through inter-dimer coupling $J_2$. (b) Two independent trillium lattices coupled via exchange couplings $J_3$ and $J_5$ and linked through inter-trillium coupling $J_4$, forming a hypertrillium lattice.	(c) Temperature dependence of the $ac$ magnetic susceptibility at 100 Hz, with the inset displaying an image of a single crystal of KSrFe$_{2}$(PO$_{4}$)$_{3}$. (d) Low-temperature magnetic specific heat of KSrFe$_{2}$(PO$_{4}$)$_{3}$ measured in zero magnetic field. The solid line corresponds to the power-law fitting.
 (e--h) Corresponding figures for K$_{2}$FeSn(PO$_{4}$)$_{3}$, in which two Fe sites are partially substituted by Sn ions. Panel (h) additionally shows the magnetic specific heat under different applied magnetic fields.  Adapted from \cite{10.1063/5.0096942} and \cite{lmsf73hn} with permission from AIP and APS, respectively, and from Ref.~\cite{ChoiPrivateComm}.	}{\label{KFSO}}.
 \end{figure*}Magnetic-field-tunable suppression of antiferromagnetic order is common in quantum magnets, though QSL states typically exist only within narrow field ranges before transitioning  to polarized or other phases. In K$_2$Ni$_2$(SO$_4$)$_3$, however, fields above 4\,T reveal a broad specific heat maximum near 1\,K that persists up to 14\,T, indicating robust short-range spin correlations rather than a fully polarized, gapped state~\cite{PhysRevLett.127.157204}. This suggests a field-induced QSL ground state emerging above $\mu_0 H = 4$\,T.\\
The nature of excitations in this field-induced QSL remains unclear, but thermodynamic signatures offer important clues. It is known that a polarized state shows low-temperature magnetic specific heat with $C_{m}$ $\sim\exp(-\Delta/T)$ behavior. In contrast, K$_2$Ni$_2$(SO$_4$)$_3$ exhibits nearly quadratic specific heat up to 14\,T, with no sign of a gapped state or conventional magnon ($\sim T^3$) behaviour~(see figure~\ref{KNSO}(e)). This raises the possibility that K$_2$Ni$_2$(SO$_4$)$_3$ harbors a QSL-like state over a broad field range of $\mu_0 H = 4$\,T to $(25$--$30)$\,T, below the anticipated polarized field~\cite{Gonzalez2024}.\\
  To probe spin dynamics and low-energy excitations in K$_2$Ni$_2$(SO$_4$)$_3$, zero-field and longitudinal-field muon spin relaxation ($\mu$SR) measurements were performed. Despite long-range magnetic order, the muon asymmetry showed no oscillations, indicating predominantly dynamic local moments~(see Ref.~\cite{PhysRevLett.127.157204}). At low-temperatures, the relaxation spectra were best modeled by a combination of a stretched exponential component (fast-relaxing, arising from unpaired spins) and a slower Lorentzian term (from the ordered background). Both relaxation rates ($\lambda_1$ and $\lambda_2$) show plateau-like behavior at low temperatures, suggesting persistent spin dynamics. Notably, the stretched exponential exponent ($\beta \approx 2$) alludes to Gaussianlike relaxation from fluctuating unpaired spins. Such relaxation behavior, also seen in the 3D QSL candidate PbCuTe$_2$O$_6$~\cite{PhysRevLett.116.107203}, is characteristic of singlet-like correlated ground states and supports the presence of spinon-like fractional excitations in K$_2$Ni$_2$(SO$_4$)$_3$~\cite{PhysRevLett.73.3306}.\\\\
 Furthermore, inelastic neutron scattering (INS) experiments by two independent groups reveal spin continuum excitations persisting even below $T_{\rm N}$. This indicates strong quantum fluctuations and QSL-like features~\cite{Gonzalez2024,PhysRevLett.131.146701}, akin to those observed in the Kitaev QSL candidate $\alpha$-RuCl$_3$~\cite{Banerjee2016}. Analysis of the INS data suggests that enhanced spin frustration originates from dominant $J_4$ and $J_5$ exchange interactions, forming a hypertrillium spin topology~(figure~\ref{KNSO}(a)). \\
 Noteworthy is that the magnetic ground state is highly sensitive to the magnitude of exchange parameters~\cite{PhysRevLett.131.146701}. When only $J_4 = J_5$ interactions are considered, a spin-liquid ground state is stabilized~\cite{Gonzalez2024}. Since both $J_4$ and $J_5$ are leading terms in K$_2$Ni$_2$(SO$_4$)$_3$, the system lies near a quantum critical point (QCP) where the ordered magnetic moment is suppressed~(figure~\ref{KNSO}(f)).

\subsection{\normalfont\textbf{ Anomalous spin dynamics in high-spin double trillium lattices}}
High-spin frustrated systems are promising hosts of CSL states, which often involves complex spin correlation patterns within a subset of spin configurations, with the number of such configurations growing rapidly as the system size increases~\cite{Balents2010}. These systems can also support exotic excitations such as fractons, and their low-energy behavior is often described by emergent gauge theories~\cite{Lozano2024}.\\
Beyond pyrochlore magnets, experimental realizations of CSLs with fractionalized excitations and emergent gauge fields remain rare. The interconnected trillium lattice offers a complementary platform for such studies. Notably, the single trillium lattice compound Na[Mn(HCOO)$_3$] ($S$ = 5/2) shows CSL-like features above its magnetic ordering temperature.
This raises an intriguing question: does coupling two trillium lattices with $S$ = 5/2 spins support a genuine CSL state, or does the hypertrillium lattice topology enhance spin fluctuations, as observed in K$_2$Ni$_2$(SO$_4$)$_3$?\\
In this context, the recently synthesized double trillium lattice compounds KSrFe$_2$(PO$_4$)$_3$ and K$_2$FeSn(PO$_4$)$_3$ (Fe$^{3+}$, $S$ = 5/2) are representative candidate materials~\cite{10.1063/5.0096942,lmsf73hn}. Both crystallize in the chiral space group $P2_1 3$, with KSrFe$_2$(PO$_4$)$_3$ showing K/Sr site mixing, and K$_2$FeSn(PO$_4$)$_3$ exhibiting Fe/Sn site disorder. While Fe ions occupy the 4$a$ site in KSrFe$_2$(PO$_4$)$_3$, they reside on the 12$b$ site in K$_2$FeSn(PO$_4$)$_3$, with a shorter inter-trillium bond in the latter~(figure~\ref{KFSO}(a-b) and (e-f). Magnetic susceptibility measurements yield CW temperatures of $-70$~K for KSrFe$_2$(PO$_4$)$_3$ and $-43$~K for K$_2$FeSn(PO$_4$)$_3$ (see Table~\ref{Table}), indicating dominant antiferromagnetic interactions. The reduced CW temperature in the Fe/Sn site-disordered system likely originates from dilution of the magnetic ions, leading to the decrease in the number of nearest-neighbor spins.
\\
DFT calculations for KSrFe$_2$(PO$_4$)$_3$ yield five antiferromagnetic couplings: $J_1 = -4.87$~K, $J_2 = -2.67$~K, $J_3 = -3.94$~K, $J_4 = -4.99$~K, and $J_5 = -5.34$~K \cite{10.1063/5.0096942} (figure~\ref{KFSO}(a,b). Unlike K$_2$Ni$_2$(SO$_4$)$_3$, where $J_1$ and $J_2$ are weak, here all $J_i$ are significant and antiferromagnetic. Notably, $J_1 \approx 0.91 J_5$, indicating that intra-dimer interactions are substantial and can perturb the spin dynamics governed by the hypertrillium topology defined by $J_4$ and $J_5$~\cite{Gonzalez2024}. It is worth noting that for the ideal hypertrillium network with $J_4 = J_5$, theoretical studies predict the absence of long-range magnetic order persisting down to $T = 0.001J_4$ even in the classical limit ($S \to \infty$). Figures~\ref{KFSO}(a,b) and (e,f) illustrate two representative spin configurations, emphasizing the important role of dimer couplings along with hypertrillium spin interactions in Fe$^{3+}$ based double trillium lattices.\\
\\
The magnetic properties of the related $S = 5/2$ system, KBaFe$_2$(PO$_4$)$_3$, were first investigated by Battle \textit{et al.} using neutron diffraction and M\"ossbauer spectroscopy~\cite{BATTLE198616}. They proposed the presence of an L-type ferrimagnetic order with $T_{\rm C}$ between 3.9 and 4.2~K, and reported strong internal magnetic fields of 516~kOe (site 1) and 494~kOe (site 2). However, the precise origin of the magnetic ordering and internal fields remains unresolved to date. In addition, recent measurements on similar compound  KSrFe$_2$(PO$_4$)$_3$ reveal an anomaly near 3.4~K, observed as a splitting between ZFC and FC magnetic susceptibility~\cite{10.1063/5.0096942}. This anomaly is also observed in the $ac$ magnetic susceptibility of single crystals, as shown in figure~\ref{KFSO}(c). Nonetheless, the magnetic specific heat (figure~\ref{KFSO}(d)) exhibits no clear signature near this temperature, suggesting the subtle nature of the transition. Notably, the low-temperature magnetic specific heat exhibits an almost quadratic dependence on temperature, suggesting the presence of nontrivial spin excitations.\\
On the other hand, the diluted compound K$_2$FeSn(PO$_4$)$_3$ (figure~\ref{KFSO}(e) and (f)) exhibits similar behavior and has been proposed to feature weak canted antiferromagnetic order driven by DM interactions~\cite{lmsf73hn}, emerging around 2~K—slightly lower than the transition observed in KSrFe$_2$(PO$_4$)$_3$~(figure~\ref{KFSO}(g)). The observed near-quadratic behavior in the specific heat has been attributed to the formation of local spin singlet (figure~\ref{KFSO}(h)). Additionally, electron spin resonance (ESR) measurements revealed a temperature-dependent linewidth $\Delta_H \propto T^{-p}$ with $p \approx 1.12$–$1.5$, indicative of low-dimensional spin correlations and raising questions about the role of the hypertrillium spin topology in high-spin double trillium lattices~\cite{lmsf73hn}.\\
To further probe the ground state, $\mu$SR experiments were performed by J. Khatua \textit{et al.} which revealed dominant dynamic spin fluctuations coexisting with weak static magnetic order in K$_2$FeSn(PO$_4$)$_3$ \cite{lmsf73hn}. The relaxation spectra exhibit a non-decoupled Gaussianlike component at low temperatures, suggestive of quasi-static internal fields potentially arising from low-energy excitations. Further theoretical investigations incorporating subleading interactions, such as DM interactions and three-spin chiral terms, may be necessary to  capture the complexity of  quantum magnetic ground states in $S = 5/2$ double trillium lattice compounds.
\\\begin{figure*}[t]
	\centering
	\includegraphics[width=\linewidth]{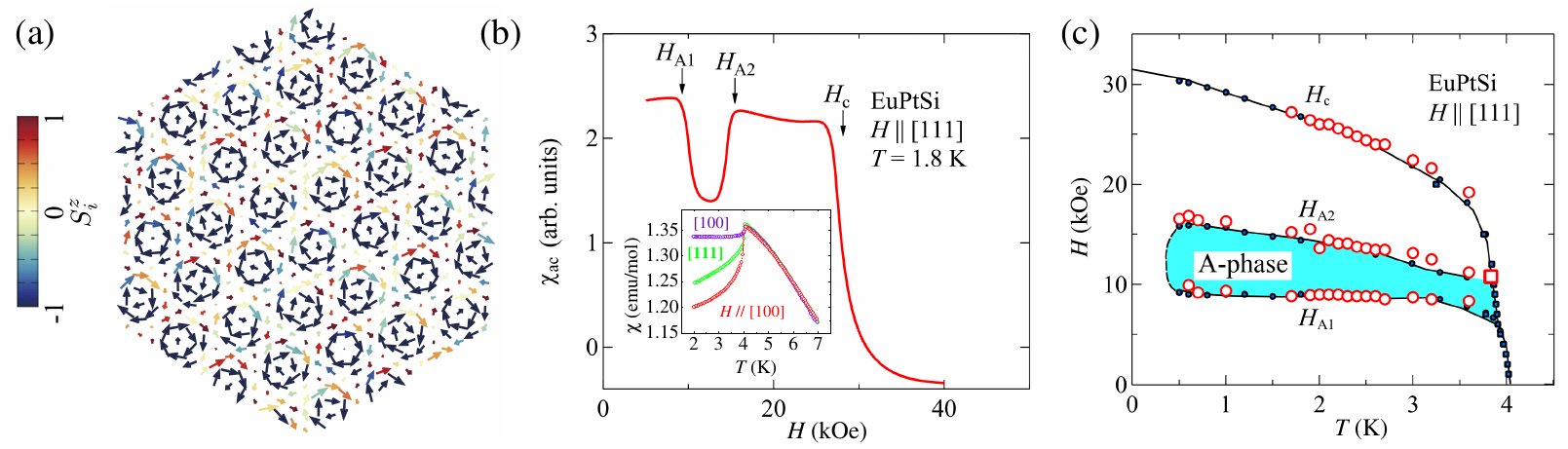}
	\caption{(a) Spin configuration in the skyrmion phase of EuPtSi along the [111] direction, based on theoretical calculations with Dzyaloshinskii–Moriya and biquadratic interactions. Arrows show the in-plane spin projection, and contours indicate the out-of-plane component.
		(b) \textit{ac} magnetic susceptibility as a function of magnetic field applied parallel to the [111] direction at $T = 1.8$~K. Two magnetic transitions occur at fields $H_{A1} = 9.2$~kOe and $H_{A2} = 13.8$~kOe, followed by saturation above $H_c = 26.6$~kOe.
		The bottom inset shows the temperature dependence of magnetic susceptibility at $H = 5$~kOe for different orientations.
		(c) Field–temperature phase diagram for a magnetic field applied along the [111] direction, where the A-phase is identified as a skyrmion phase. Adapted from \cite{doi:10.7566/JPSJ.90.073705} and \cite{nuki2020} with permission from JPSJ.	}{\label{EPT}}.
\end{figure*}

\subsection{\normalfont\textbf{Magnetic skyrmion}}\label{SKYR}
The interplay between magnetic order and topology in spin systems engenders exotic phenomena, closely tied to the symmetry and topology of spin configurations. Discovering new topological spin arrangements remains a major focus in condensed matter research~\cite{Paddison2024}. Among these, skyrmions are well-known topological spin textures observed in chiral magnets~\cite{Tokura2021}.
Skymions are swirl-like, particle-like structures with remarkable stability due to their topological nature (figure~\ref{EPT}(a)). Originally proposed in high-energy physics to describe baryons~\cite{SKYRME1962556}, skyrmions now play a key role in understanding topology in real space and hold promise for spintronic applications due to their nanoscale size, efficient current-driven motion, and topological protection~\cite{Tokura2021}
\\
The first experimental observation of skyrmions was made in the trillium lattice-based helimagnet MnSi~\cite{science1166767}, where they appeared within a narrow temperature window at the boundary of the long-range helical phase under an applied magnetic field. In MnSi, the skyrmion phase emerges from a delicate balance between ferromagnetic exchange and DM interactions. These DM interactions, which originate from spin–orbit coupling in non-centrosymmetric crystals, are crucial for stabilizing the topological spin textures~\cite{science1166767}.\\
Singularly, skyrmions can also emerge in centrosymmetric metals due to competing long-range Ruderman–Kittel–Kasuya–Yosida (RKKY) and biquadratic exchange interactions~\cite{PhysRevLett.108.017206,KHATUA20231}. Skyrmions in frustrated chiral magnets have  gained interest, with a notable example being the trillium lattice compound EuPtSi, which crystallizes in the non-centrosymmetric space group $P2_1 3$, similar to MnSi~\cite{Kakihana2018}. In EuPtSi, Eu$^{3+}$ ions (4$f^7$; $L = 0$, $S = 7/2$) constitute a 3D network of corner-sharing triangles~\cite{Kaneko2019, PhysRevB.96.014401}. Magnetism in this system arises from RKKY interactions, where localized 4$f$ moments couple indirectly via spin-polarized conduction electrons.\\
Magnetic susceptibility shows a sharp drop at $T_{\rm N} = 4$\,K, signaling a phase transition with noticeable anisotropy below this temperature~\cite{ADROJA1990375,nuki2020,Kakihana2017} (see inset of figure~\ref{EPT}(b)). This transition is proposed to be of first-order, supported by a steep resistivity drop and a pronounced specific heat peak~\cite{Kakihana2017}. At high temperatures and low fields, EuPtSi has a positive CW temperature ($\sim$ 4\,K), suggesting dominant ferromagnetic interactions. This  has led to its consideration as a candidate for spin-ice physics on the trillium lattice within a ferromagnetic Ising framework~\cite{PhysRevB.82.014410}. However, a crossover to antiferromagnetic correlations occurs around 1.2\,T, as shown by $^{29}$Si NMR studies  by Higa \textit{et al.}~\cite{PhysRevB.104.045145}. A partial entropy release  ($\sim$ $0.5\,R\ln 8$) at $T_{\rm N}$, along with with critical slowing down of Eu spin fluctuations from $\mu$SR and NMR measurements, points to competing exchange interactions above $T_{\rm N}$~\cite{PhysRevB.96.014401,Homma2019}. A combined effort of theoretical and further experimental studies is required to reveal whether the observed short-range correlations above $T_N$ indeed originate from chiral spin fluctuations. \\\\
Moreover, the trillium compound EuPtSi hosts field-induced magnetic phases that are highly sensitive to both the strength and direction of the applied magnetic field. When a field is applied along the [111] direction, there appear two metamagnetic transitions at $H_{\rm A1} = 9.2$ kOe and $H_{\rm A2} = 13.8$ kOe, followed by the onset of a field-polarized state at $H_c$. As shown in figure~\ref{EPT}(b), the $ac$ magnetic susceptibility displays a dip between $H_{\rm A1}$ and $H_{\rm A2}$, suggesting the emergence of an intermediate magnetic phase, referred to as the A-phase. The corresponding field–temperature phase diagram (figure~\ref{EPT}(c)) reveals that the A-phase spans a temperature range from 0.5 K to 3.6 K for the $H \parallel$ [111] direction. In EuPtSi, the A-phase has been identified as a skyrmion phase through various experimental techniques, including Hall resistivity, magnetostriction, and magnetoresistance measurements~\cite{nuki2020,Kakihana2017}. \\ \\ Unlike MnSi, the skyrmion phase in EuPtSi is highly anisotropic, with a much smaller size of approximately 18~\AA, about one order of magnitude smaller than that in MnSi. Neutron scattering in EuPtSi reveals magnetic ordering at $\mathbf{q}1 = (0.2, 0.3, 0)$ below 0.3 K, with a first-order commensurate–incommensurate transition near $T^{*}_{\rm N} \sim 2.5$ K.  On the other-hand, in the A-phase under 1.2 T $\parallel$ [111], magnetic peaks form a hexagonal pattern with ordering vector $\mathbf{q}_{\rm A} \simeq (\pm0.09, \pm0.20, \mp0.28)$, indicative of skyrmion lattice formation \cite{Kaneko2019}. Further resonant x-ray diffraction measurements unveil that the A-phase of EuPtSi hosts a triple-$\mathbf{q}$ magnetic structure with all components sharing the same helicity, each perpendicular to its respective $\mathbf{q}$ vector~\cite{PhysRevB.109.174437}. This uniform helicity, consistent with the low-field helimagnetic phase, highlights the role of chiral antisymmetric DM exchange interactions in stabilizing the triangular skyrmion lattice~\cite{PhysRevB.109.174437}. Recent theoretical studies suggest that two key magnetic interactions—the DM interaction and the biquadratic interaction mediated by itinerant electrons—play a crucial role in stabilizing skyrmions in EuPtSi~\cite{doi:10.7566/JPSJ.90.073705}.\\ Beyond itinerant magnets, oxide-based trillium lattice compound Cu$_2$OSeO$_3$ also exhibits the skyrmion phases, characterized by a double-$\mathbf{q}$ ordering stabilized by the anisotropic DM interactions~\cite{Chacon2018}. In such chiral system, the competition between DM interactions and ferromagnetic exchange typically result in a spiral ground state at zero field, while an external magnetic field induces the skyrmion phases \cite{DZYALOSHINSKY1958241}.\\  The absence of inversion symmetry in certain crystals gives rise to intriguing spin–momentum entanglement, leading to antisymmetric spin splitting in the electronic band structure.  For example, this effect manifests as Weyl-type spin splitting in chiral systems \cite{PhysRevB.94.235117} and Rashba-type splitting in polar symmetric structures~\cite{Ishizaka2011}. Such spin–momentum coupling also underlies various emergent phenomena, including the spin Hall effect~\cite{spinhall}.  The recent discovery of lanthanide-based magnets with chiral trillium lattices has opened new possibilities for exploring topological magnetic phases. It remains an intriguing question whether the interplay between weak DM interactions and strong antiferromagnetic Heisenberg exchange can stabilize a skyrmion phase in the zero-field ground state of such systems. Along this line, the recently studied trillium compound K$_2$CrTi(PO$_4$)$_3$ appears to be a promising candidate \cite{PhysRevB.109.184432}, exhibiting a magnetic transition at 4.3 K and signatures of spin fluctuations above this temperature—reminiscent of behavior observed in EuPtSi.

\section{Synthesis challenges and chiral effect detection}
\subsection{Challenges in material growth}
\textcolor{black}{To date, synthesis of trillium materials remains limited to a few well-characterized examples, with conventional flux-growth or solid-state methods often producing small crystals or powders and generating significant antisite disorder.\\
Na[Mn(HCOO)$_3$] has been realized primarily in polycrystalline form, although small single crystals have been reported using slow-diffusion synthesis methods~\cite{PhysRevLett.128.177201}. These small crystals, while adequate for structural determination, are unsuitable for advanced spectroscopic or neutron scattering and thermal Hall experiments that require large, untwinned, single-domain samples. Developing reliable synthesis routes capable of producing large, high-quality single crystals  will be essential to enable definitive measurements of chiral spin fluctuations.\\
An alternative class of candidate materials resides within the large family of trillium-lattice langbeinite compounds (see Table~\ref{jk}), many of which remain essentially unexplored regarding their magnetic ground-state properties. Though millimeter-sized single crystals of certain langbeinites have been obtained via conventional flux- and melt-growth techniques~\cite{lmsf73hn,PhysRevLett.127.157204}, further development of crystal growth strategies is needed to increase crystal size and quality. This need is particularly acute because many phosphate-based langbeinites suffer from significant antisite disorder, which complicates disentangling intrinsic magnetic behavior from disorder-induced effects.
Recent studies have revealed that some langbeinite compounds undergo structural phase transitions at low temperatures while retaining their chiral space group at room temperature~\cite{cavatri,magar2025proximate}. These findings underscore the importance of thoroughly examining thermal and structural stability.\\
Adding to the complexity is the structural motif of langbeinites, which involves two interpenetrating trillium lattices forming a so-called hypertrillium lattice constructed of corner-sharing tetrahedra. To realize and investigate exotic magnetic phases in these double-trillium systems, it is essential to synthesize langbeinite crystals free from cation disorder, as such disorder can mask intrinsic chiral behaviors and hinder emergent phenomena. Alternatively, selecting combinations of magnetic $M''^{3+}$ and non-magnetic $N''^{4+}$ ions that avoid site mixing can effectively isolate a single trillium framework from the double trillium lattice, simplifying the structural complexity.}

\subsection{Probing chirality}
\textcolor{black}{To probe chiral effects in trillium-lattice magnets and disentangle the roles of chirality and geometric frustration in their dynamic spin fluctuations, advanced spectroscopic and transport techniques should be combined.\\
Magnetic circular dichroism (MCD) offers direct insight into chirality by exploiting the interplay of structural chirality and magnetization in non-centrosymmetric crystals. The direction-dependent optical response reverses upon flipping either the light propagation vector or magnetization~\cite{Bordcs2012}. Importantly, MCD can also probe the quantum geometry of the electronic states through light-matter coupling, providing access to fundamental properties like Berry curvature and quantum metric that underlie topological and chiral phases. However, reliable MCD measurements require strict control over sample quality—specifically, ensuring enantiopurity and detwinning, since multiple enantiomers can cancel the MCD signal, and controlling magnetic domains because the signal scales with the projection of light propagation on magnetization~\cite{PhysRevB.87.014421,Tokura2018}.\\
Thermal conductivity, particularly thermal Hall measurements, provide a complementary probe of chiral spin correlations. While phonons and magnons are charge-neutral and do not experience a Lorentz force, systems with strong spin–orbit coupling, Berry curvature, or DMI can exhibit analogous thermal Hall effects~\cite{Czajka2023}. Chiral spin liquids are predicted to show quantized thermal Hall conductivity, a definitive signature that distinguishes them from cooperative paramagnets~\cite{PhysRevB.110.214430,Lozano2024,PhysRevLett.119.127204,ZHANG20241}. Achieving and conclusively demonstrating this quantization remains an experimental priority. Separating phononic from magnetic contributions in thermal transport is challenging but crucial; recent studies on systems such as YMnO$_3$ have uncovered thermal transport signatures consistent with chiral spin fluctuations~\cite{Kim2024}.\\
Alongside thermal transport, polarized neutron scattering is another powerful means to probe the antisymmetric part of the dynamic susceptibility, allowing direct detection of chiral spin fluctuations not only in ordered phases but also above ordering temperatures where short-range chiral correlations persist~\cite{PLAKHTY1999259,Simonet2012}. Studies on the trillium-lattice magnet MnSi have revealed intrinsic chiral paramagnetic fluctuations derived from DMI that survive across phase transitions~\cite{ROESSLI2004124}. Similarly, polarized diffuse neutron scattering on the non-centrosymmetric compound YBaCo$_3$FeO$_7$ has detected pronounced chirality within strongly interacting disordered spin states~\cite{PhysRevX.12.021029}. A key future step is to apply these MCD, thermal transport, and polarized neutron scattering techniques to promising trillium-lattice magnetsits to unravel chiral spin correlations and its interplay with frustration.}

\section{Beyond the known chiral magnets and topological phenomena}
Chirality in materials science is  a compelling notion due to its unique ability to interact with light, electrons, and other particles in asymmetric ways, enabling novel functionalities in optics~\cite{Hallett2022}, electronics~\cite{YooPark}, spintronics~\cite{PhysRevLett.124.136404,Tokura2021},  and biomedicine~\cite{pharmaceutics14091951}. Ongoing research continues to unlock its potential, driven by advances in synthesis, characterization, and theoretical modeling~\cite{GOBEL20211}. In this review, we have highlighted the rich spin topologies and exotic many-body phenomena found in the trillium class of chiral magnets, which crystallize in the non-centrosymmetric space group $P$2$_1$3 (see Table.\ref{table}). While some compounds feature a single trillium lattice, members of the  langbeinite family exhibit more complex spin structures arising from two interpenetrating trillium sublattices, which are weakly coupled by exchange interactions. To deepen our understanding of such phenomena and to uncover new topological states governed by the interplay of lattice chirality and spin interactions it is crucial to examine structurally simpler analogues.\\
A critical first step in this direction is identifying of a compound’s space group to determine its crystallographic chirality. Among the 230 space groups, 65 Sohncke space groups lack inversion and mirror symmetry, allowing for the possibility of chiral properties in crystalline solids~\cite{sohncke1879entwickelung}. For example, the chiral space group $P$3$_{1}$21, one of the Sohncke groups, is realized in compounds such as DyFe$_{3}$(PO$_{4}$)$_{3}$, which hosts quadrupole helix chirality. Similarly, the chiral space group $P6_{3}$ is found in BaCoSiO$_{4}$, known to host toroidal magnetic moments~\cite{Ding2021}.\\
Looking ahead, the discovery of low-spin trillium-lattice compounds with chiral space group symmetry and enhanced quantum fluctuations may pave the way for realizing long-sought topological magnetic phases. Among the proposed phases, the chiral QSL stands out as a particularly exotic state—it lacks conventional long-range magnetic order yet spontaneously breaks both time-reversal and parity symmetries~\cite{PhysRevB.39.11413}. Conceptually, the chiral QSL can be viewed as the magnetic analog of the fractional quantum Hall effect, characterized by topological order and fractionalized excitations~\cite{PhysRevLett.59.2095}. Interestingly, it has been proposed that 2D quantum magnets which break reflection and time‑reversal symmetries can harbor topologically non‑trivial ground states classified by Chern numbers, mirroring the quantized Hall conductance seen in the quantum Hall effect~\cite{RevModPhys.58.519}.  Theoretical investigations on various two-dimensional spin systems employing the spin Hamiltonian~\cite{PhysRevB.109.125146,Banerjee2023}

\begin{equation}
\mathcal{H} =
\sum_{\langle i,j \rangle} J_{ij}\,\mathbf{S}_i \cdot \mathbf{S}_j \;+\;
\sum_{\langle i,j,k \rangle_{\triangle}} J_{ijk}^{\triangle}\,
\mathbf{S}_i \cdot \big(\mathbf{S}_j \times \mathbf{S}_k\big),
\end{equation}

where the first term represents the Heisenberg exchange and the second term describes chiral three-spin interactions on triangular plaquettes, have predicted a variety of exotic magnetic phases, including chiral QSL states.
 It remains an open and exciting question whether such a chiral QSL state can be realized in chiral magnets with trillium geometry. As a starting point, the already promising QSL candidate K$_2$Ni$_{2}$(SO$_4$)$_3$ could be investigated to help constrain and map out the phase diagram, shedding light on the conditions under which field-induced, chiral QSL behavior emerges~\cite{PhysRevLett.127.157204}. Ultimately, polarized INS and thermal Hall measurements will be essential to definitively confirm its existence.
\\
Moreover, for high-spin pyrochlore systems, theoretical studies have predicted the existence of chiral CSL states, further emphasizing the universality of chiral spin-interaction-driven disordered state in frustrated magnets~\cite{Lozano2024,Liu2024}. Given the inherent DM interaction in chiral magnets, these systems are also prominent candidates for materializing  spiral spin liquid ground states—where spins fluctuate collectively among degenerate spiral configurations, maintaining a disordered yet correlated state~\cite{Yao2021}. Recent theoretical studies on honeycomb lattice have considered the spin Hamiltonian~\cite{PhysRevB.111.174435}
\begin{equation}
\mathcal{H} =
\sum_{\langle i,j \rangle} J_{ij}\,\mathbf{S}_i \cdot \mathbf{S}_j
+ \sum_{\langle i,j \rangle} \mathbf{D}_{ij} \cdot (\mathbf{S}_i \times \mathbf{S}_j)
- B \sum_{i} S_i^z ,
\end{equation}
where the first and second terms represent the Heisenberg exchange and DM interaction, respectively, and the third term corresponds to a Zeeman coupling due to an external magnetic field \( B \) applied along the \( z \)-axis. This model has been shown to host a spiral spin liquid state at elevated temperatures, which transitions into  magnetic skyrmion lattices at lower temperatures. Exploring analogous physics in the trillium lattice may offer new insights. In this direction, the observation of short-range spin correlations above $T_{\rm N}$ in the $S=5/2$ trillium lattice K$_{2}$FeSn(PO$_{4}$)$_{3}$ suggests it could be a promising candidate for realizing a spiral spin liquid state~\cite{PhysRevB.109.184432}. Furthermore, related spin models have also been proposed to stabilize quantum skyrmion phases in low-spin systems, offering new possibilities for exploring chiral topological magnetism in chiral magnets~\cite{liscak2025}.\\
Beyond the exotic magnetism arising from the absence of inversion symmetry in chiral magnets, such systems also exhibit intriguing topological phenomena, including the topological phonon Hall effect~\cite{PhysRevLett.105.225901} and chiral phonons~\cite{Zhang2025}. The latter—quanta of lattice vibrations analogous to circularly polarized electromagnetic waves—appear as circularly polarized vibrational modes carrying intrinsic angular momentum~\cite{PhysRevLett.112.085503,scienceaar,PhysRevLett.121.175301}. This emerging field is gaining attention for significant implications for quantum information science~\cite{PhysRevLett.121.175301,PhysRevLett.132.056302}, superconductivity~\cite{Grissonnanche2020}, and phononic engineering \cite{zhang2025chirali,Uchida2008}.
Chiral phonons in various materials have been experimentally detected using techniques such as inelastic X-ray scattering~\cite{PhysRevLett.121.035302}, Raman spectroscopy, and pump-probe spectroscopy~\cite{Ishito2023,Zhang2023}, and supported by density functional theory calculations~\cite{Ueda2023}. Notably, phonon Zeeman effect measurements on Cu$_{2}$OSeO$_{3}$  lead us to identify chiral phonons carrying large effective magnetic moments due to their coupling to magnons in the magnetically ordered state~\cite{ChoiPrivateComm}.
Interestingly, recent DFT calculations on the trillium lattice compound EuPtSi reveal the existence of chiral phonons in this non-centrosymmetric system~\cite{PhysRevB.111.165132}. This finding suggests that newly discovered langbeinite-based trillium lattice compounds may also provide promising platforms for investigating chiral quasiparticle dynamics. One intriguing possibility is to manipulate magnetism by leveraging angular momentum transfer between chiral phonons and spins, enabled by shining chiral materials with circularly polarized light.

\section{Conclusion}
Quantum fluctuations are rare in 3D systems, yet frustrated magnets like the trillium lattice-based compound provide notable exceptions, sustaining exotic ground states despite their higher connectivity. The trillium lattice, with its intrinsic chiral symmetry and geometrically frustrated exchange interactions, offers a compelling platform for realizing emergent quantum phenomena in 3D magnets. It serves as a complementary setting for exploring exotic spin textures and topological states, advancing both fundamental understanding and potential applications in spintronics. Beyond conventional symmetric Heisenberg interactions, the trillium lattice accommodates subleading antisymmetric interactions—such as Dzyaloshinskii–Moriya and other chiral couplings—that open new research avenues into topological magnetism.\\
While several intriguing properties, including noncollinear magnetic ordering and magnetic skyrmions, have been observed in trillium-based compounds, a complete understanding of more exotic phenomena—such as chiral spin liquids, spin fractons, and phonon Hall effects in chiral magnets—remains an open challenge. In this review, we have outlined the theoretical developments and highlighted experimental realizations of trillium compounds, emphasizing the need for further investigation into unexplored magnetic phases and unconventional excitations. Future progress will rely on discovering new materials, refining theoretical frameworks, and employing advanced experimental probes to uncover the full potential of trillium-based chiral magnets.

\section{References}

\bibliographystyle{iopart-num}
\bibliography{Trillium}
\end{document}